\documentclass[a4paper, 11pt]{article}
\usepackage[top=2.5cm, bottom=2.5cm, left=2.5cm, right=2.5cm]{geometry}
\usepackage{graphicx}
\usepackage{epstopdf, epsfig}
\usepackage{amsmath}
\usepackage{amssymb}
\usepackage{comment}
\usepackage{natbib}
\usepackage{booktabs}
\usepackage[table]{xcolor}
\usepackage{soul}
\usepackage{bm}
\usepackage{subcaption}
\usepackage{caption, subcaption, floatrow}
\usepackage{pdflscape}
\usepackage{multicol}

\usepackage{myglossaries}

\newcommand{\MMG}[1]{#1}

\title{\textbf{Particle-resolved  simulations of settling particles: \\ A methodology for long time-integration intervals }}

\author{M. Moriche$^{1}$, M. García-Villalba$^{1}$ and M. Uhlmann$^{2}$ \\[2ex]
 $^{1}$ Institute of Fluid Mechanics and Heat Transfer\\ TU Wien, Vienna,  Austria \\[1ex]
$^{2}$ Institute for Water and Environment, \\Karlsruhe Institute of Technology, Karlsruhe, Germany \\[1ex]
\small Corresponding author: manuel.moriche@tuwien.ac.at   \\
{\it Accepted in Acta Mechanica}
}
\date{}

\newcommand{\eqsMov}{\eqref{eq:fno:mom}--\eqref{eq:pno:ang}}
\newcommand{\eqsLab}{\eqref{eq:fng:mom}--\eqref{eq:png:ang}}

\begin{document}

\maketitle

\begin{abstract}

We present a methodology for simulating dilute suspensions of particles settling
under gravity, with the main purpose of overcoming limitations of triply
periodic configurations, mainly the strong vertical correlation that hinders the
study of cluster dynamics.
The current approach removes vertical periodicity and employs a moving reference
frame, enabling efficient simulations of both single- and
many-particle cases.
We illustrate the method with two examples of increasing complexity: a single
particle in the steady vertical regime, and a many-particle case at a parametric
point where collective effects were previously observed and recovered here.
A converged, free-of-corrections time interval of approximately $600\,D/U_g$
is simulated in the many-particle case, representing the first simulation of this
kind to date.
New physical insights can be explored thanks to this new configuration,
for example the effect of still fluid on the first layer of particles encountered
by the fluid, or the turbulent character of the flow after a swarm of
particles has passed by.
Finally, the method only requires parameter tuning, allowing implementation
within existing solvers without changes to their core formulation:
\MMG{for a standard configuration with an imposed free stream velocity at
the inlet, only the input velocity (or the viscosity of the fluid) and the
time step need to be updated.}
\end{abstract}

\section{Introduction}

Particle-laden flows, in which solid particles interact with a 
surrounding fluid, play critical roles in both natural and industrial 
systems \citep{subramaniam:2023,marchioli2025}. 
Such flows are challenging because both experiments and simulations are hard to carry out.
Experimental characterization requires high-resolution three-dimensional 
tracking of particle positions and orientations in opaque suspensions \citep{villafane2025}.
Numerical methods using particle-resolved direct numerical simulations 
(PR-DNS) face also a complex task. 
They must capture detailed hydrodynamic interactions at the scale of 
individual particles, while also accounting for large-scale collective 
behavior \citep{uhlmann:2023,garciavillalba:2025}. 
This multiscale challenge requires enormous computational resources.
Even state-of-the-art simulations
in dilute systems
typically handle only $O(10^4)$ particles at moderate Reynolds numbers, 
limiting statistical significance.

One particular problem of interest in this field involves
gravity-driven, particle-laden flows in the dilute regime, where collective 
effects significantly modify the hydrodynamic behavior compared to 
isolated particles. 
These collective effects manifest 
themselves
as spatial 
heterogeneities in particle concentration, with experimental 
observations revealing regions of higher particle concentration (clusters) 
and lower (voids) relative to the mixture's average 
concentration \citep{huisman:2016}. 
The onset of clustering fundamentally alters the flow 
dynamics, resulting in enhanced settling velocities, increased 
collision frequencies, and amplified fluid perturbations, among other 
modified transport properties.

Numerical simulations of such systems are typically conducted in 
triply periodic configurations, analyzing particle behavior in the 
absence of walls or other geometric constraints beyond the fluid-particle
interfaces.
This computational setup has yielded valuable insights into clustering phenomena for spherical particles \citep{kajishima:2002,uhlmann:2014a,fornari:2016}, cubes \citep{seyed-ahmadi:2021}, oblate spheroids \citep{fornari:2018, moriche:2023} and prolate spheroids \citep{lu:2023,jiang:2024}. 
However, the triply periodic approach presents inherent limitations 
for studying cluster dynamics. 
As clusters grow to sizes comparable to the
computational domain, the results become strongly correlated along 
the vertical direction.
This constrains our ability to analyze cluster evolution and stability 
beyond their initial formation stages. 
%
%
To address these limitations, this work 
introduces a methodology that enables an 
efficient usage of inflow–outflow 
conditions, thereby removing the need for 
vertical periodicity.
This non-homogeneous vertical setup 
overcomes
the challenge that the terminal settling 
velocity of the particle ensemble is not known a priori.
The method uses an iterative approach, 
adding corrections to determine 
this a-priori-unknown velocity, 
or what is equivalent,
the  Reynolds number of the particle ensemble.
In this way, we enable long time-integration 
intervals in reasonably small
computational domains and without 
the need to consider non-inertial effects
in the governing equations.

The article is organized as follows. First, in \S \ref{sec:problem} we 
describe in detail the problem under study, including the governing 
equations, the relevant non-dimensional parameters, and the definition of 
the reference frames that will be used throughout the paper. 
Then, in \S \ref{sec:meth} we present the proposed methodology, where we 
explain how the iterative correction of the Reynolds number is implemented, 
together with the mapping operators and the details regarding the numerical 
integration.
In \S \ref{sec:example_single_particle} and \S \ref{sec:example_many}, two 
representative examples are shown, first for the case of a single particle 
in the steady vertical regime, and later for the case of many particles 
where collective effects appear and the performance of the method is 
evaluated. 
Finally, the main conclusions and remarks are summarized in 
\S \ref{sec:conclusions}, where we also discuss 
the potential of the approach 
to be applied to other problems and the possible directions for future work.

%
%
%
%

\section{Problem description}
\label{sec:problem}

\begin{figure} 
\makebox[\textwidth][c]{ 
\includegraphics[scale=0.8]{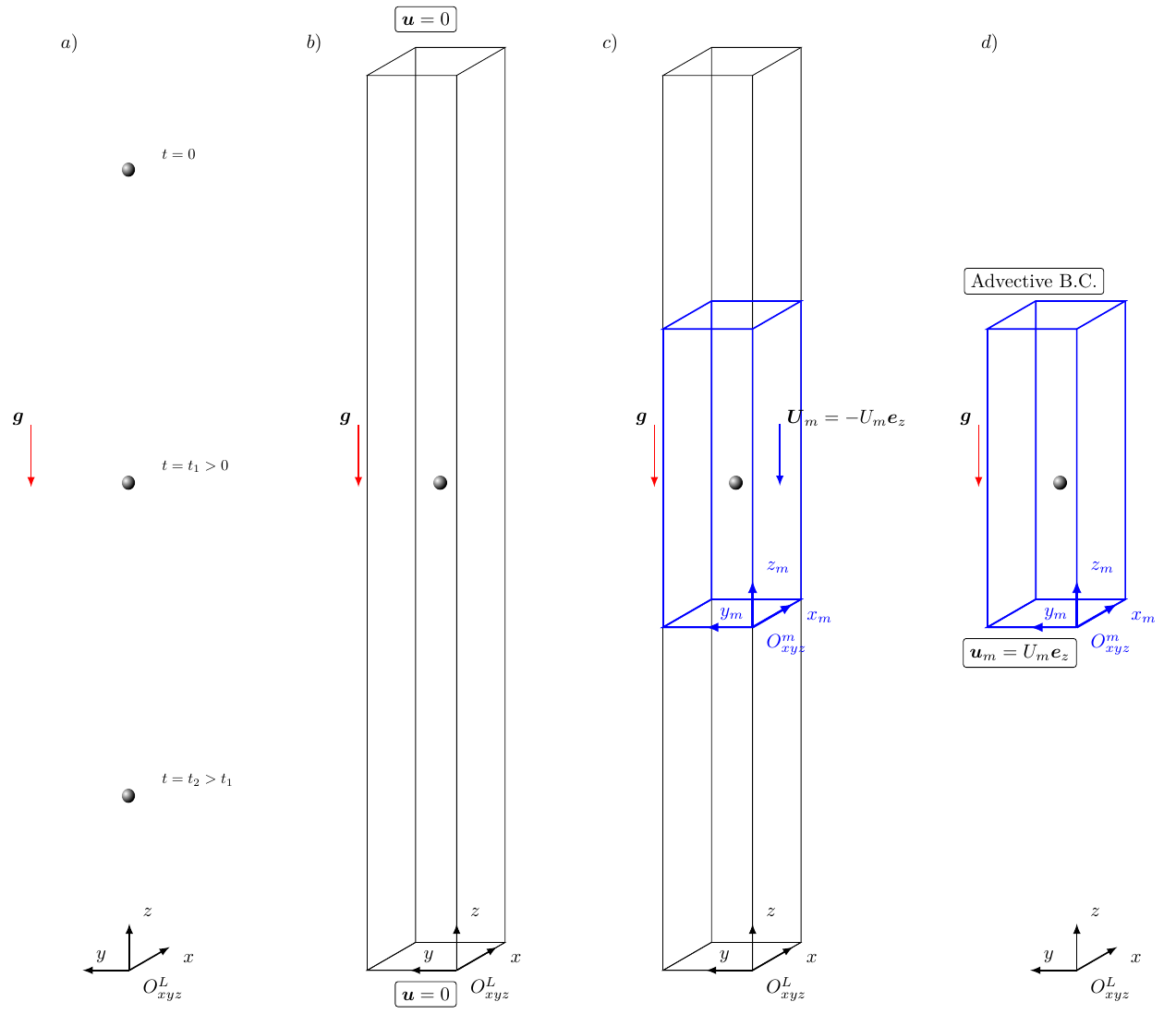} 
}
\caption{a) Sketch of the particles settling in the unbounded domain  and b)
representation of a computational domain in the laboratory frame,
\gls{set:Olab}.  c) Sketch of the relation of the laboratory $(\gls{set:Olab})$
and moving $(\gls{set:Otra})$ frames.
d) Representation of the computational domain in the moving frame,
\gls{set:Otra}.
The gravitational acceleration \gls{vg} is included in all the panels and the
boundary conditions in the bottom and top boundaries are indicated in panels b
and d.
\label{fig:problem}} 
\end{figure}

The problem under study is the settling of particles under the effect of
gravity in an unbounded domain.
The fluid is assumed to be Newtonian with density \gls{f:rho} and kinematic
viscosity \gls{f:nu}, and the particles are assumed to be rigid spheres with
uniform density \gls{p:rho}, volume \gls{p:vol} and moment of inertia per unit
mass \gls{p:spi}.
Note that the extension to non-spherical particles does not affect the
methodology, but it is not considered here to simplify notation.
We further assume the flow to be incompressible.
Figure \ref{fig:problem}a shows a sketch of the problem including the
gravitational acceleration vector, \gls{vg}, the particles at three different
instants (assuming $\gls{p:rho}>\gls{f:rho}$), and the laboratory frame,
\gls{set:Olab}.
The laboratory frame is defined as the reference frame in which the fluid is at
rest in the absence of particles.
The vertical direction is defined as the direction opposite to gravitational
acceleration, with unit vector $\gls{ez} = -\gls{vg}/\left|\gls{vg}\right|$.
In the following we assume particles to be heavy and, for simplicity of
notation, the governing equations are discussed for the case of a single
particle; the change in formulation for lighter-than-the-fluid particles
and the extension to an ensemble of particles being straightforward.
%
%


In the laboratory frame \gls{set:Olab} the governing equations for the fluid
velocity, \gls{f:u}, and pressure, \gls{f:p}, are the Navier-Stokes equations
for an incompressible flow
\begin{align} 
\ppt{\gls{f:u}} + \left(\gls{f:u}\cdot\nabla\right) \gls{f:u} &=
   -\cfrac{\nabla\gls{f:p}}{\gls{f:rho}}
   + \gls{f:nu} \nabla^2\gls{f:u} \,,
  \label{eq:f:mom}\\
\nabla \cdot \gls{f:u} &= 0\,, \label{eq:f:cont}
\end{align}
and for the particle velocity, $\gls{p:u}=\left(\gls{p:u_x},\gls{p:u_y},
\gls{p:u_z}\right)$, and angular velocity, \gls{p:o}, the Newton-Euler
equations for a rigid body
\begin{align}
\gls{p:rho}\gls{p:vol}\ddt{\gls{p:u}} &= 
 \int_S \gls{f:stress} \cdot \gls{n} \wrt \gls{dA}
 + \gls{p:vol}\left(\gls{p:rho}-\gls{f:rho}\right)\gls{vg}
 \,,  \label{eq:p:lin}\\
\gls{p:rho}\gls{p:vol}\ddt{(\gls{p:spi}\gls{p:o})} &=
 \int_S \gls{r} \times \left(\gls{f:stress} \cdot \gls{n} \right)\wrt \gls{dA}
 \,, \label{eq:p:ang}
\end{align}
where $S$ represents the surface of the particle, \gls{n} is a unit vector
normal to the surface of the particle pointing towards the fluid and \gls{r} is
a position vector with respect to the center of gravity.
The fluid stress tensor, \gls{f:stress}, is defined in terms of the fluid
pressure and velocity as $\gls{f:stress} = -\gls{f:p}\gls{eye} + \gls{f:rhonu}
\left(\nabla \gls{f:u}+\nabla\gls{f:u}^T\right)$, where \gls{eye} is the
identity matrix.
\MMG{
Please note that the term involving gravity is explicitly added in
\eqref{eq:p:lin} because the gravity term has been eliminated from the fluid
momentum equation \eqref{eq:f:mom}, as is customarily done by subtracting the
hydrostatic part from the pressure field.
%
%
Note also that for an ensemble of particles, a collision term has to be added
to the right hand side of \eqref{eq:p:lin}. 
For a discussion of the various collisions models, we refer to the literature
\citep{glowinski1999,kempe2012,kidanemariam2014,costa2015}.
}
%

Dimensional analysis shows that, in addition to 
the solid volume fraction, the problem is governed by two non-dimensional
parameters.
Among the possible combinations, we select the density ratio between the
particle and the fluid, \gls{rhor}, and the Galileo number, \gls{Ga}, defined as
\begin{align}
\gls{rhor} & = \cfrac{\gls{p:rho}}{\gls{f:rho}}        \,, \\
\gls{Ga}   &=  \cfrac{\gls{Ug}\gls{p:D}}{\gls{f:nu}}   \,,
\end{align}
where $\gls{p:D}$ is the diameter and
$\gls{Ug}=\sqrt{\left(\gls{rhor}-1 \right)\left|\gls{vg}\right|\gls{p:D}}$ is a
gravitationally-scaled velocity.
%
%

If we choose $\gls{f:rho}$, $\gls{p:D}$ and  $\gls{Ug}$ as reference
quantitites, we can rewrite equations \eqref{eq:f:mom}-\eqref{eq:p:ang} in
non-dimensional form as
\begin{align} 
\pptg{\left(\gls{fng:u}\right)}
   + \left(\gls{fng:u}\cdot\gls{nabla_n}\right) \gls{fng:u} &=
   -\gls{nabla_n}\gls{fng:p}
   + \frac{1}{\gls{Ga}}\gls{nabla_n}^2\gls{fng:u} \,,
   \label{eq:fng:mom}\\
   \gls{nabla_n} \cdot \gls{fng:u} &= 0\,,
   \label{eq:fng:cont} \\
   \gls{rhor}\gls{pn:vol}\ddtg{\left(\gls{png:u}\right)} &= 
    \int_S \left(-\gls{fng:p}\gls{eye}
   + \frac{1}{\gls{Ga}}
      \left(\gls{nabla_n}\gls{fng:u}+\left(\gls{nabla_n}\gls{fng:u}\right)^T\right)\right)
    \cdot \gls{n} \wrt \gls{dA_n}
    - \gls{pn:vol}\gls{ez} \,,
   \label{eq:png:lin}\\
   \gls{rhor}\gls{pn:vol}\ddtg{\left(\gls{pn:spi}\gls{png:o}\right)} &=
    \int_S \gls{r_n} \times  \left(-\gls{fng:p}\gls{eye}
 +\frac{1}{\gls{Ga}} \left(\gls{nabla_n}\gls{fng:u}+\left(\gls{nabla_n}\gls{fng:u}\right)^T
      \right)\right)
   \cdot \gls{n} \wrt \gls{dA_n} \,,
   \label{eq:png:ang}
\end{align}
where the particle parameters ($\gls{pn:vol} = \gls{p:vol}/\gls{p:D;cb}$
and $\gls{pn:spi}=\gls{p:spi}/\gls{p:D;sq}$), the differential element of area
($\wrt\gls{dA_n}=\wrt\gls{dA}/D^2$) and the position vector ($\gls{r_n}=%
\gls{r}/D$) are also expressed in non-dimensional form.
We also introduce the non-dimensional nabla operator normalized with the
particle diameter $\gls{nabla_n}=\left(\ppxn{},\ppyn{},\ppzn{}\right)$.
Given the values of 
\gls{rhor} and \gls{Ga},
equations \eqref{eq:fng:mom}-\eqref{eq:png:ang} can be solved numerically with
proper boundary conditions.
For a Cartesian domain like the one shown in figure \ref{fig:problem}b, the
velocity at the bottom and top boundaries is set to zero ($\gls{f:u}=0$).
In this work, periodicity is applied as a lateral boundary condition to 
approximate an unbounded domain; nevertheless, the method we present is not 
limited to periodic boundaries in the lateral directions.

In the following sections, we will rewrite the equations with respect to moving
reference frames. 
To maintain clarity, we express non-dimensional quantities for both fluid and
particles as ratios of dimensional quantities to their corresponding reference
values. 
While this approach makes equations \eqref{eq:fng:mom}--\eqref{eq:png:ang} 
appear more complex, it reduces the number of variable definitions and prevents
reader confusion when we present the proposed methodology. 
For quantities whose non-dimensional form remains constant across all reference
frames in this work, such as particle geometric properties, we denote the
non-dimensional version by adding a tilde to the symbol.

\subsection{The need of a moving reference frame}
\label{sec:problem/moving}

Figure \ref{fig:problem}b shows a sketch of the computational domain used to
integrate equations \eqref{eq:fng:mom}--\eqref{eq:png:ang} numerically.
Any combination of a finite Galileo number ($\gls{Ga}>0$) and a
non-neutrally-buoyant particle ($\gls{rhor}>1$) will produce a
relative motion of the particle with respect to the fluid.
As illustrated in figure \ref{fig:problem}a,
after transients have been discarded,
heavy particles will settle with an average
terminal velocity.
%
In order to accommodate the initial transient and a sufficiently long
time interval of interest, the domain must be elongated in the vertical direction.
%
%
As an alternative, it is possible to use 
a non-periodic domain (in the vertical 
direction) in conjunction with
a moving frame, \gls{set:Otra}, which
travels at a constant velocity $\gls{vUo} = -\gls{Uo}\gls{ez}$ (with $\gls{Uo}=%
\left|\gls{vUo}\right|$), with respect to the laboratory frame
\citep{uhlmann:2014b,moriche:2021,catalan2024}.
This is sketched in 
figure
\ref{fig:problem}c.
The velocity of the fluid and the particle with respect to the moving frame are
defined as $ \gls{fo:u}=\gls{f:u}-\gls{vUo}$ and $\gls{po:u}=%
\gls{p:u}-\gls{vUo}$, respectively.
Since the velocity \gls{vUo} is constant, we can write the Navier-Stokes
equations for the pressure and fluid velocity with respect to the moving frame,
\gls{f:p} and \gls{fo:u}, respectively, as
\begin{align} 
\ppt{\gls{fo:u}} + \left(\gls{fo:u}\cdot\nabla\right) \gls{fo:u} &=
   -\frac{\nabla\gls{f:p}}{\gls{f:rho}}
   + \gls{f:nu} \nabla^2\gls{fo:u} \,,
  \label{eq:fo:mom}\\
\nabla \cdot \gls{fo:u} &= 0\,.
\label{eq:fo:cont}
\end{align}
Similarly, we can write the Newton-Euler equations for the particle relative
motion
\begin{align}
\gls{p:rho}\gls{p:vol}\ddt{\gls{po:u}} &= 
 \int_S \gls{f:stress} \cdot \gls{n} \wrt \gls{dA}
 + \gls{p:vol}\left(\gls{p:rho}-\gls{f:rho}\right)\gls{vg}
 \,,  \label{eq:po:lin} \\
\gls{p:rho}\gls{p:vol}\ddt{\left(\gls{p:spi}\gls{p:o}\right)} &=
 \int_S \gls{r} \times \left(\gls{f:stress} \cdot \gls{n} \right)\wrt \gls{dA}
 \,. \label{eq:po:ang}
\end{align}
Note that the pressure \gls{f:p}, the angular velocity \gls{p:o}, and the fluid
stresses are invariant to Galilean transformation, therefore we do not need to
define new symbols for these quantities in the moving frame.
Equation \eqref{eq:po:ang} is the same as equation \eqref{eq:p:ang} and it is
repeated for readability.
The change in reference frame implies that the boundary condition at the bottom
boundary of the computational domain is $\gls{fo:u} = \gls{Uo}\gls{ez}$ and a 
typical advective or stress-free condition type should be imposed at the top
boundary (see figure \ref{fig:problem}d).

If we choose $\gls{f:rho}$, $\gls{p:D}$ and  $\gls{Uo}$ as reference
quantities, we can rewrite equations \eqref{eq:fo:mom}-\eqref{eq:po:ang} in
non-dimensional form as
\begin{align} 
\ppto{\left(\gls{fno:u}\right)} + \left(\gls{fno:u}\cdot\gls{nabla_n}\right) \gls{fno:u} &=
   -\gls{nabla_n}\gls{fno:p}
   + \frac{1}{\gls{Reo}}\gls{nabla_n}^2\gls{fno:u} \,,
  \label{eq:fno:mom}\\
\gls{nabla_n} \cdot \gls{fno:u} &= 0\,, \label{eq:fno:cont} \\
\gls{rhor}\gls{pn:vol}\ddto{\left(\gls{pno:u}\right)} &= 
 \int_S \left(-\gls{fno:p}\gls{eye}
+ \frac{1}{\gls{Reo}} \left(\gls{nabla_n}\gls{fno:u}+\left(\gls{nabla_n}\gls{fno:u}\right)^T\right)\right)
 \cdot \gls{n} \wrt \gls{dA_n}
 - \gls{pn:vol}\left(\frac{\gls{Ga}}{\gls{Reo}}\right)^2\gls{ez}
 \,,  \label{eq:pno:lin}\\
\gls{rhor}\gls{pn:vol}\ddto{\left(\gls{pn:spi}\gls{pno:o}\right)} &=
 \int_S \gls{r_n} \times  \left(-\gls{fno:p}\gls{eye}
+ \frac{1}{\gls{Reo}} \left(\gls{nabla_n}\gls{fno:u}+\left(\gls{nabla_n}\gls{fno:u}\right)^T\right)\right)
\cdot \gls{n} \wrt \gls{dA_n}
 \,. \label{eq:pno:ang}
\end{align}
where we have introduced the Reynolds number based on the velocity of the
moving frame with respect to the laboratory frame \gls{Uo},
\begin{equation} \label{eq:Reo}
\gls{Reo} = \frac{\gls{Uo}\gls{p:D}}{\gls{f:nu}}  \,.
\end{equation}

Given the values of 
\gls{rhor}, \gls{Ga}
and \gls{Reo}, we could solve numerically equations \eqsMov{}.
The aim of integrating equations \eqsMov{} instead of 
\eqsLab{} is twofold: i) to enable long time-integration intervals in relatively
small computational domains, as sketched in Figure \ref{fig:problem}c, leading
to an affordable computational cost, and ii) to use existing numerical tools
without having to implement ad hoc terms to account for non-inertial effects.
In order to achieve these goals it is essential to properly select the value of
\gls{Reo}, which should correspond to the particle Reynolds number based on the
terminal velocity of the particle (or particle ensemble),
$\langle\gls{p:u_z}\rangle_t$,
\begin{equation}\label{eq:Rep}
\gls{Rep}  = \frac{-\langle\gls{p:u_z}\rangle_t\gls{p:D}}{\gls{f:nu}} \,,
\end{equation}
 where the operator $\langle
\cdot\rangle_t$ represents time averaging.

\begin{figure} 
\makebox[\textwidth][c]{ 
\includegraphics[scale=1.0]{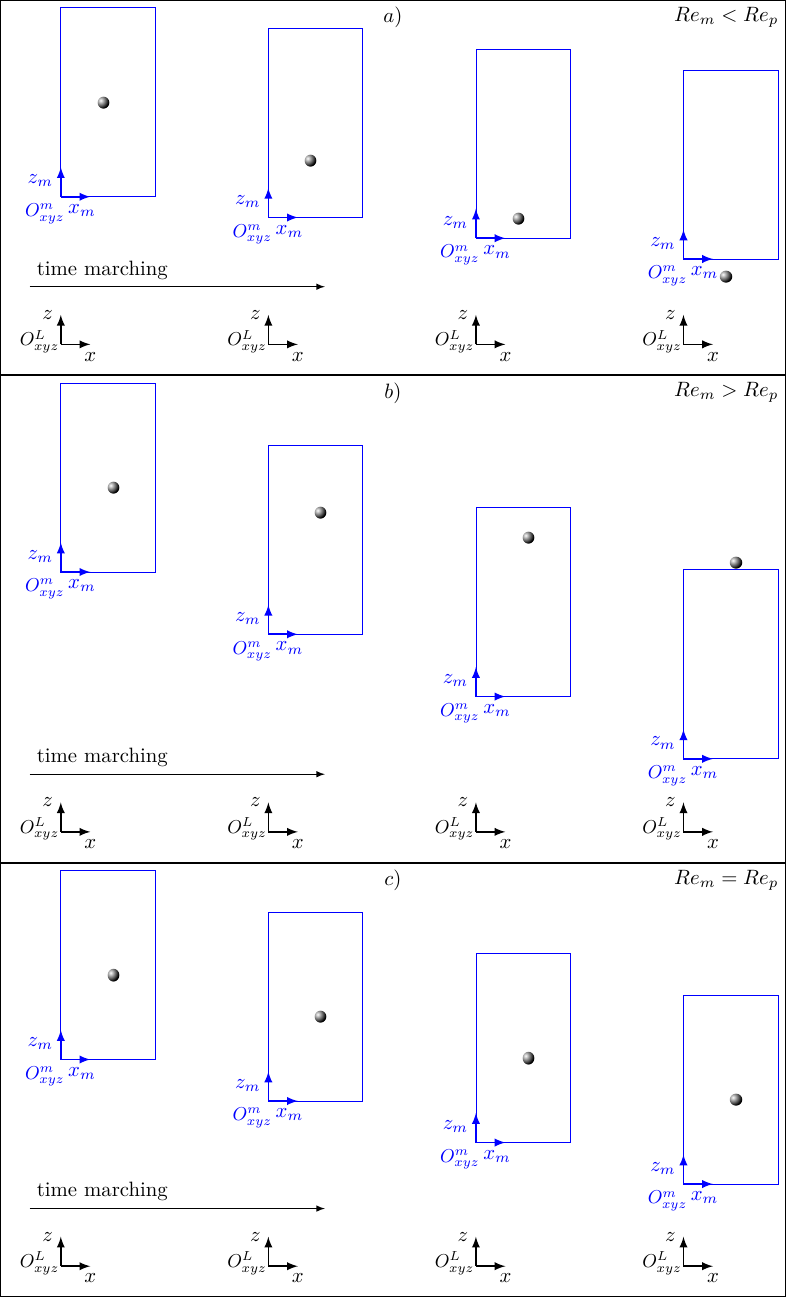} 
}
\caption{Sketch of the possible scenarios for values of \gls{Reo} a) lower
, b) higher or c) equal to the  particle Reynolds number \gls{Rep}
\label{fig:particles_leaving}} 
\end{figure} 

The problem with this approach is that the particle Reynolds number, \gls{Rep},
is a result of the calculation, not known a priori.
As a consequence, the precise value of $\gls{Reo}$ to achieve the goal of long
integration time intervals is not known at the beginning of the simulation.
Indeed, we would like to have $\gls{Reo} = \gls{Rep}$, which would imply
\begin{align}\label{eq:goal}
\gls{Reo} = \gls{Rep}  
    \Leftrightarrow  \gls{Uo} =  -\langle \gls{p:u_z} \rangle_t
    \Leftrightarrow  \left\langle \gls{pno:u_z} \right\rangle_t = 0 \,,
\end{align}
resulting in no average drift of the particle in the vertical direction in the
moving reference frame.
Figure \ref{fig:particles_leaving} shows the three possible
scenarios regarding the value of \gls{Reo} with respect to \gls{Rep}:
\begin{itemize}
\item $\gls{Reo} < \gls{Rep}$: Particles drift towards the bottom of the domain
      and eventually are influenced  by the presence of the bottom boundary.
\item $\gls{Reo} > \gls{Rep}$: Particles drift towards the top of the domain and
      eventually are influenced  by the presence of the top boundary.
\item $\gls{Reo} \approx \gls{Rep}$: Particles stay in the computational domain
      for a long time interval.
\end{itemize}
For a single particle, the scenarios discussed above comprise all possible
cases.
However, for an ensemble of particles, additional considerations may arise, 
particularly regarding the homogeneity of the mixture. 
For example, particles may drift apart from each other in certain situations. 
Even when the particle Reynolds number of the ensemble is properly captured,
individual particles might drift and become influenced by the presence of top
and/or bottom boundaries.
The potential negative influence of these individual particles drifting away
will depend on the total number of particles and should be evaluated with care.

In the following section we present a method to iteratively correct the value of
\gls{Reo} enabling long time-integration intervals in reasonably small
computational domains (as sketched in  figure \ref{fig:particles_leaving}c), and
using a numerical tool in which equations \eqsMov{} are discretized, without 
the need of implementing ad hoc terms to account for non-inertial effects.

\section{Methodology}
\label{sec:meth}

Here we propose an iterative method to obtain a sequence of estimates of
\gls{Reo}, denoted \gls{Reo_s}, being $s$ the iteration index, that, if
successful, converges towards the particle (or particle ensemble) Reynolds
number.
This is done by letting the system evolve during consecutive, relatively short
time intervals and using the resulting evolution to correct the next iterate.
It will be shown that, even if the initial guess is a rough estimate, the
method is robust and converges efficiently.
Please note that in this work, we use `iteration' to refer to a time interval 
during which the system evolves using the same \gls{Reo_s}. 
Each iteration consists of many individual time steps.


Let us first outline the methodology before proceeding with the technical
details.
We start with an initial condition, such as the fluid and particles at rest in
the laboratory frame, and specify an initial value for the Reynolds number
\gls{Reo_0}, which can be estimated using simple order-of-magnitude methods or
any other suitable approach.
Assuming that we have already performed $s-1$ iterations, then the following
steps are to be followed for the next iteration $s$:
\begin{enumerate}
   \item We change the non-dimensional state variables at the end of the
   previous iteration, \gls{tB_sm1}, from the laboratory frame \gls{set:Olab} to
   the moving reference frame \gls{set:Otra_s}. This is done using a simple
   mapping operation to be discussed below.
   \item We then integrate equations \eqsMov{} in a time interval $\left[\gls{tA_s},
   \gls{tB_s}\right]$ with an estimated value of the  Reynolds number
   \gls{Reo_s}. The length of the time interval can be determined during run
   time. For example, when a predefined particle vertical displacement threshold
   is met, the time interval ends. During integration one must accumulate the
   complete time evolution of a reduced set of particle-related statistics
   (compared to the full set of state variables), which will be later used to
   estimate the Reynolds number of the next iteration, \gls{Reo_sp1}.
   \item The non-dimensional state variables are now transferred from the
   moving reference frame \gls{set:Otra_s} to the laboratory reference frame
   \gls{set:Olab}, by using the inverse mapping operation of step 1. This is
   done in two substeps:
   \begin{itemize}
      \item The reduced set of statistics is mapped for the whole time interval 
      $\left[\gls{tA_s},\gls{tB_s}\right]$.
      \item The full set of state variables is mapped only for the last time instant
      \gls{tB_s} to initialize the next iteration.
   \end{itemize}
   \item We then concatenate the reduced set of statistics obtained in this
   $s$-th interval  to that of the previous intervals. In this way, at this
   stage the complete time evolution in the interval $[0,t_B^{(s)}]$ is
   available for the statistics needed to compute \gls{Reo_sp1}.
   \item Using the complete time evolution of the reduced set of statistics we
   can correct the Reynolds number and determine its value
   for the next iteration, \gls{Reo_sp1}. This involves a control/optimization
   strategy which is problem dependent and it is discussed in its own section
   below.
   \item We can now return to step 1 for the next iteration
\end{enumerate}
Please note that the choice of retaining intermediate fluid and/or particle
states in step 2 is left to the user; we show here the steps using the minimum
amount of storage resources for the algorithm to function.

\subsection{Elements of the algorithm}
\label{sec:meth/els}

\subsubsection{Set of state variables}
\label{sec:meth/els/state}

We define the two sets of non-dimensional variables each of which fully defines
the state of the system:
\begin{equation} \label{eq:state_lab}
\gls{lab:state}(\vec{x},t;\gls{Ga},\gls{rhor}) :=  \left\{
   \gls{fng:u}, \gls{fng:p}, \gls{png:u}, \gls{png:o}   \right\} \,,
\end{equation}
associated with the laboratory frame \gls{set:Olab}, and
\begin{equation} \label{eq:state_tra}
\gls{tra:state_s}(\vec{x},t;\gls{Ga},\gls{rhor},\gls{Reo_s}) :=  \left\{
   \gls{fno_s:u}, \gls{fno_s:p}, \gls{pno_s:u}, \gls{pno_s:o}   \right\} \,,
\end{equation}
with the $s$-th moving frame \gls{set:Otra_s}.
Ideally, one would like to integrate the state vector in the laboratory frame
shown in \eqref{eq:state_lab} since it only depends on the governing parameters
of the problem, \gls{Ga} and \gls{rhor}.
However, in order to integrate the solution for long time intervals, we rely on
integrating equations \eqsMov{} with $\gls{Reo}=\gls{Reo_s}$ to obtain the state
vector \gls{tra:state_s}.
Therefore, the main purpose of working with \gls{tra:state_s} is numerical
integration.
%

\subsubsection{Mapping functions to change the reference frame}
\label{sec:meth/els/map}

In this section we define the mapping functions \gls{Mlm} and \gls{Mml} to
change the reference frame from the laboratory to the moving frame and vice
versa, respectively,
\begin{align}
\gls{tra:state} &= \gls{Mlm}\left( \gls{lab:state}; \gls{gam} \right) \,, \\ 
\gls{lab:state} &= \gls{Mml}\left( \gls{tra:state}; \gls{gam} \right) \,,
\end{align}
where we introduce the ratio between the velocity of the moving frame and the
gravitationally-scaled velocity
\begin{equation}
\gls{RoG} = \frac{\gls{Uo}}{\gls{Ug}}
          = \frac{\gls{Reo}}{\gls{Ga}}   \,.
\end{equation}
Assuming that the moving frame travels 
with respect to the laboratory frame
along the vertical direction $\gls{vUo} =
-\gls{Uo}\gls{ez}$ (see figure \ref{fig:problem}c) the resultant expressions are
\begin{alignat}{2}
\gls{tra:state} &= \gls{Mlm}\left(\gls{lab:state};\gls{RoG}\right)\,\Rightarrow && \,
\begin{cases}
\gls{fno:u} & = \gls{fng:u}\,\gls{GoR} + \gls{ez} \\
\gls{fno:p} & = \gls{fng:p}\,\gls{GoR;sq}\\
\gls{pno:u} & = \gls{png:u}\,\gls{GoR} + \gls{ez} \\
\gls{pno:o}& =  \gls{png:o}\,\gls{GoR} 
\end{cases} \,, \\
\gls{lab:state} &= \gls{Mml}\left(\gls{tra:state};\gls{RoG}\right)\,\Rightarrow && \,
\begin{cases}
\gls{fng:u} & = \left( \gls{fno:u} - \gls{ez}\right)\gls{RoG} \\
\gls{fng:p} & = \gls{fno:p}\,\gls{RoG;sq}\\
\gls{png:u} & =\left(\gls{pno:u} - \gls{ez}\right)\gls{RoG}\\
\gls{png:o}& =  \gls{pno:o}\,\gls{RoG} 
\end{cases} \,.
\end{alignat}
Please note that we have dropped the iteration index ($s$) for clarity.

\subsubsection{Integration}
\label{sec:meth/els/int}

As discussed above, numerical integration is performed in the moving reference
frame.
Figure \ref{fig:problem}d shows the computational setup in which the problem is
integrated: an inflow/outflow configuration.
More specifically, the bottom boundary condition is of Dirichlet type, 
imposing a uniform free stream of velocity $\gls{Uo}\gls{ez}$.
We apply an advective boundary condition at the top to minimize computational
domain influence, though other approaches achieving this goal are possible.
Similar freedom of choice applies for the lateral boundary conditions.
Particularly, we assume the domain to be periodic in the lateral directions
($x$ and $y$ according to figure \ref{fig:problem}d), but the algorithm has no
restrictions in the usage of other type of boundary conditions.
This flexibility is an advantage: 
unknown effects of certain lateral boundary conditions 
on particle settling pose no additional challenge.

The integration time interval of the $s$-th iteration is $[\gls{tA_s},\gls{tB_s}
]$, where the starting point is the end of the previous iteration $\gls{tA_s}=
\gls{tB_sm1}$.
There are two issues to consider when defining the duration of the time
integration interval: 
first, when $\gls{Reo_s}\ne \gls{Reo_sm1}$, the flow variables after restart 
present a small unphysical discontinuity that is damped after a few time steps.
This is due to the change in reference frame and the presence of the non-linear
term in the momentum equation. 
%
%
Since our main interest is in reaching a statistically stationary state after
having applied corrections and then keeping \gls{Reo} constant, and we are not
primarily interested in the transient state, we disregard this issue in the
following.
Nevertheless, the length of the intervals should be sufficiently long to
accommodate the damping of the small discontinuity.
Based on our experience, a minimum of $10$ time steps is a good rule of thumb.
Second, rough estimates of \gls{Reo}, although handled by the method, may
result in moderately large vertical drifts inside the computational domain.
Therefore, the first set of iterations should be relatively short to avoid large
displacements of the particle(s) in the moving frame.
%
%
This requirement can be relaxed if, during runtime, the simulation is designed to stop when a 
specified condition is violated, for example, when the particle’s vertical position falls 
below or rises above predefined thresholds.

The other aspect to consider for numerical integration is the selection of the
time step. 
In this work we use a fixed time step normalized with the gravitationally-scaled
velocity, $\gls{ng:dt;flat}$, which means that the time step is updated with
every correction of \gls{Reo_s}.
Therefore, the non-dimensional time step used to integrate numerically equations
\eqref{eq:fng:mom}--\eqref{eq:pno:ang} is computed as follows
\begin{equation}\label{eq:meth/dt}
\gls{no:dt_s} = \gls{RoG_s}\gls{ng:dt}   \,.
\end{equation}
%
The relationship between the 
time normalized with the gravitationally-scaled velocity,
\gls{ng:t;flat}, and the time normalized with the velocity of the moving
frame, \gls{no:t;flat}, is somewhat more elaborated than equation
\eqref{eq:meth/dt}, and is discussed in Appendix~\ref{app:time_zbot}.

\subsubsection{Correction of Reynolds number \gls{Reo}}
\label{sec:meth/els/correct}

In this section we describe how to correct the value of the  Reynolds
number between successive iterations.
%
This method component is highly dependent on the specific problem and user preferences.
Therefore, we present here the simplest expression to update the ratio
$\gls{RoG}=\gls{Reo}/\gls{Ga}$, which for a fixed Galileo number is equivalent
to obtain an expression for $\gls{Reo}=\gls{RoG}\,\gls{Ga}$, but somehow more
convenient since in this work we are presenting the mapping operators as
functions of \gls{RoG}.
Starting from $\gls{p:u_z} = \gls{po:u_z} - \gls{Uo}$, after simple
algebra, we obtain the expression
\begin{equation}\label{eq:meth/els/corr/RoG_inst}
\gls{RoG} = \cfrac{\gls{png:u_z;flat}}{\gls{pno:u_z;flat}-1} \,,
\end{equation}
which can be averaged in time, leading to
\begin{equation}\label{eq:meth/els/corr/RoG_avg}
\gls{RoG} = \cfrac{\left\langle\gls{png:u_z;flat}\right\rangle_t}{%
\left\langle\gls{pno:u_z;flat}\right\rangle_t-1} 
\,.
\end{equation}
%
Since the objective is to maintain the particle at an approximately constant
vertical position
relative to the moving reference frame, we seek
$\left\langle\gls{po:u_z}\right\rangle_t=0$.
From this condition, we can obtain
an expression to update \gls{RoG} for the next iteration
\begin{equation}\label{eq:meth/els/corr/RoG_update}
\gls{RoG_sp1} = -\left\langle\gls{png:u_z}\right\rangle_t \,,
\end{equation}
and consequently $\gls{Reo_sp1} = \gls{RoG_sp1}\gls{Ga}$.

To illustrate the algorithm's basic functioning, we consider an artificially generated test 
case mimicking single-particle settling behavior: a particle starting from rest undergoes 
constant acceleration during a transient period of duration \gls{tc} before reaching terminal 
velocity. 
Therefore, the vertical velocity relative to the laboratory frame as a function of time is
given by
%
\begin{equation}\label{eq:example_piecewise/fake}
\gls{png:u_z} =                
\begin{cases} 
-A\cfrac{t}{t_c}    \,,\, \text{if}  & 0 \le t < t_c  \\[6pt]
-A                     \,,\, \text{if}  & t_c\le t
\end{cases}
\,,
\end{equation}
where $A>0$ is a known constant.

For the iterative procedure we use fixed-duration ``integration'' intervals 
(iterations) of $\gls{Tint} = t_c/4$, and averaging intervals of the
same duration $\gls{Tavg} = \gls{Tint}$.
Therefore, the averaged value of \gls{png:u_z;flat} obtained in the $s$-th
iteration is
\begin{equation}\label{eq:meth/els/corr/int}
\left\langle \gls{png:u_z} \right\rangle_t^{(s)} = \cfrac{1}{\gls{tB_s}-\gls{tA_s}} %
   \int_{\gls{tA_s}}^{\gls{tB_s}} \gls{png:u_z}(\tau) \wrt{} \tau \,,
\end{equation}
where $\gls{tB_s}-\gls{tA_s} = \gls{Tavg} = \gls{Tint}$.
The term ``integration'' is marked by quotes because, in this particular example,
we do not integrate numerically, but use the information given in equation
\eqref{eq:example_piecewise/fake} (in the laboratory frame) and construct the 
``integrated'' solution in the $s$-th moving frame with the relation
\begin{alignat}{2} \label{eq:meth/els/corr/example_function}
\gls{pno_s:u_z} & = \gls{png:u_z}\,\gls{GoR_s} + 1           \,.  
\end{alignat}
%
%
In a realistic case, this procedure is reversed because the settling velocity 
\gls{png:u_z;flat} 
is not known a priori. One first integrates numerically in the $s$-th moving 
frame to obtain the particle-related statistics, in this case 
\gls{pno_s:u_z;flat}, and then applies the appropriate mapping functions to obtain 
\gls{png:u_z;flat}.
In the following, we will present the procedure as if it would be done in a realistic
case, using the term ``integration''.

\begin{figure} 
\makebox[\textwidth][c]{ 
\includegraphics[scale=0.8]{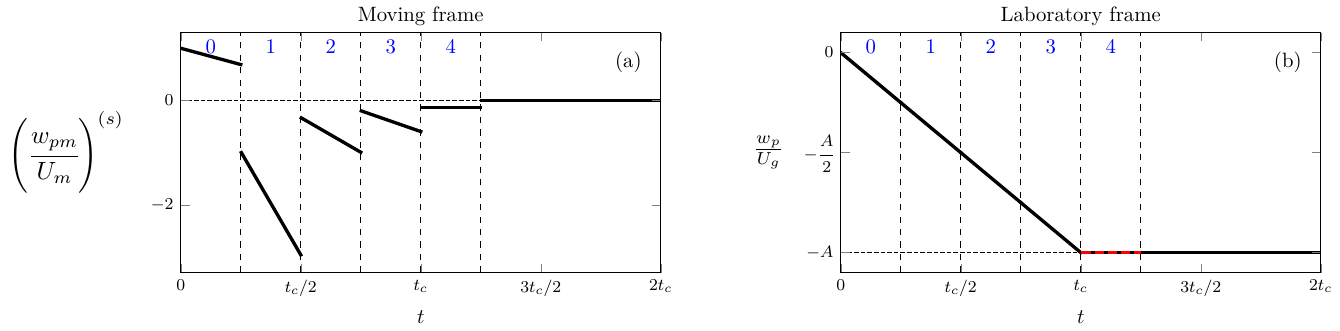} 
}
\caption{%
Evolution of the solution of the illustrative example in the a) moving 
and b) laboratory frame.
In panel b the portion of signal used to correct \gls{RoG} after the last
iteration ($s=4$) is shown with a red, dashed line.
Iterations are identified by the vertical dashed lines, with the iteration index
in blue on top of each panel.
\label{fig:correction_example}} 
\end{figure}

Figure \ref{fig:correction_example} shows the evolution of the iterative
procedure.
The problem starts from rest ($\gls{png:u_z;flat}=0 \Leftrightarrow 
\gls{pno:u_z;flat}=1$) and we initialize \gls{RoG} with an underestimated
value $\gls{RoG_0} =0.8\,A$ (simple algebra shows that the converged value of
\gls{RoG} is $A$).
After the first iteration, the particle is still at the early stages of the
transient, and therefore the averaged value of the settling velocity (see figure
\ref{fig:correction_example}b) is small compared to the terminal velocity.
We update the value of \gls{RoG} according to equation
\eqref{eq:meth/els/corr/RoG_update} and obtain $\gls{RoG_1}=0.13\,A$.
Because during the first iteration the particle is accelerating downwards from
rest, the averaged value of \gls{png:u_z;flat} used to compute \gls{RoG_1}
underestimates the instantaneous velocity at the end of the interval, therefore,
the particle has a negative initial velocity in the moving frame of the second
iteration, $\gls{pno_1:u_z;flat}(\gls{tA_1})<0$.
Furthermore, it keeps on accelerating downwards.
As commented above, this behavior deviates from our target
($\langle\gls{pno:u_z;flat}\rangle_t=0$), but it only exposes the lack of
knowledge of the solution, which is overcome by using relatively short
integration intervals, especially at the early stages.
In the consecutive iterations, we obtain $\gls{RoG_2}=0.38A$, $\gls{RoG_3}=
0.63A$ and $\gls{RoG_4}=0.88A$, which shows how the algorithm progressively
reduces the difference between \gls{RoG_s} and its target $\gls{RoG}=A$.
This can be seen in figure \ref{fig:correction_example}a, where the
vertical downward motion of the particle is smaller and smaller as the
iterative procedure advances.

The success of the algorithm starts to be visible when the transient comes to an
end.
This can be seen in figure \ref{fig:correction_example}b, where the selected
signal to compute the average of \gls{png:u_z;flat} is free from the influence
of the transient.
This results in $\gls{RoG_5}=A$, which, because of the simplicity of the current
example, is exactly our target value and results in
$\left\langle\gls{po:u_z}\right\rangle_t=0$.
From this point on, we can conclude that the iterative process is ended and the
rest of the problem is simply evolved without further modifying the value of
\gls{RoG}.

\begin{landscape}
\begin{figure}[p]
\makebox[\textwidth][c]{ 
\includegraphics[scale=1.0]{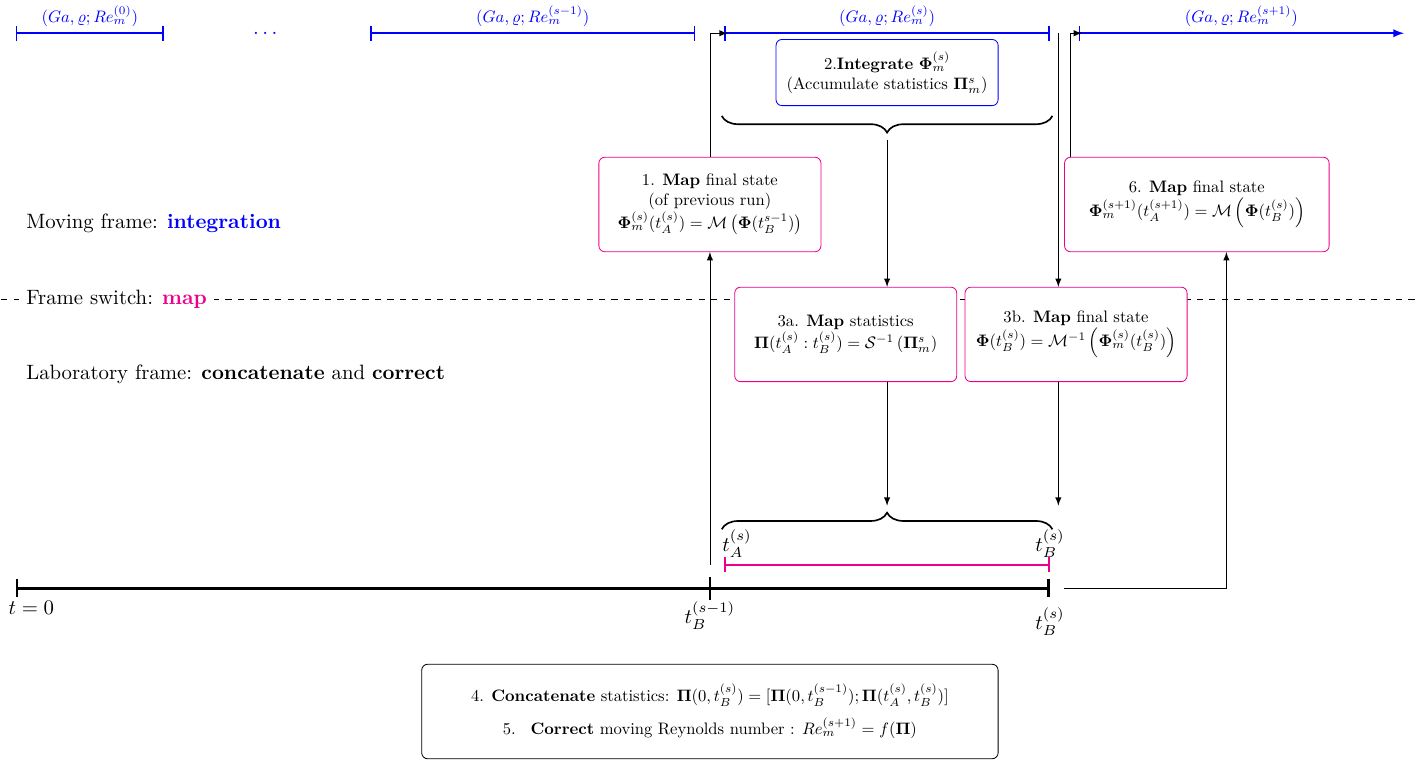} 
}
\caption{Diagram of the workflow of the algorithm 
\label{fig:algorithm_graphic}} 
\end{figure} 
\end{landscape}

Please note that, although this seems an oversimplified example, it does 
include the main features of realistic cases, especially the initial transient
behavior.
The main differences with respect to a realistic case is that more, and (probably
longer) integration intervals are needed after the transient has ended in order
to ensure that: i) the statistically stationary regime is fully attained, and
ii) the averaged values involved in updating \gls{RoG} are robust enough.
To conclude the section, we follow the notation from \S~\ref{sec:meth/els/state} and
introduce two vectors of non-dimensional quantities
that contain the reduced set of particle-related statistics used by
the correction algorithm: \gls{lab:stats},
associated with the laboratory frame, and \gls{tra:stats}, with the moving
frame.
%
%
We also define two mapping functions to transform between reference frames: \gls{Slm} maps 
from the laboratory to the moving frame, and \gls{Sml} maps from the moving to the laboratory 
frame.
In the example discussed above, this would have been
\begin{alignat}{3}
\label{eq:meth/example_stats_a}
\gls{tra:stats} & = \left\{\gls{pno:u_z}\right\}
               && = \gls{Slm}\left(\gls{lab:stats};\gls{RoG}\right)
               && = \gls{png:u_z}\,\gls{GoR} + 1 \,, 
    \\
\label{eq:meth/example_stats_b}
\gls{lab:stats} & = \left\{\gls{png:u_z}\right\}
               && = \gls{Sml}\left(\gls{tra:state};\gls{RoG}\right)
               && = \left(\gls{pno:u_z} - 1\right)\gls{RoG} \,.
\end{alignat}
The user should include in  \gls{lab:stats} (and \gls{tra:stats}) the quantities
needed for the correction algorithm employed to successfully work. 
This will become clearer in the following sections, 
where we present examples with single and multiple particles.
%
%
%
%
%
Finally, the whole methodology  is summarized in the diagram shown in 
figure \ref{fig:algorithm_graphic}.

%

\section{Settling of a single particle}
\label{sec:example_single_particle}

In this section we show the performance of the method for a single sphere
settling under gravity in the steady vertical regime.
We set the Galileo number and the density ratio to $\gls{Ga}=121$ and
$\gls{rhor}=1.5$, based on the works of \cite{jenny:2004} and
\cite{uhlmann:2014a}.
Following \cite{uhlmann:2014b} we use a cuboidal computational domain of
dimensions $[16/3 \times 16/3 \times 16] \gls{p:D;cb}$ and a spatial resolution
of $\gls{p:D}/\gls{dx}=24$ in all directions, resulting in a number of grid cells of
$[128\times128\times384]$.
We impose an inflow boundary condition with constant velocity $\gls{Uo}\gls{ez}$
at the inlet (bottom boundary), an advective condition at the outlet (top
boundary), and periodicity in the lateral directions, as sketched in
figure~\ref{fig:problem}a.

Because of the simplicity of the steady vertical regime, and the fact
that we are dealing only with one particle, the correction algorithm is very similar
to the example shown in
\S~\ref{sec:meth/els/correct}.
Thus, the reduced sets of statistics are $\gls{lab:stats}=\left\{
\gls{png:u_z;flat}\right\}$ and $\gls{tra:stats}=\left\{\gls{pno:u_z;flat}
\right\}$, with the corresponding mapping functions already presented in
equations \eqref{eq:meth/example_stats_a} and \eqref{eq:meth/example_stats_b}.
We use a time step of $\gls{ng:dt;flat}=8\cdot10^{-3}$ and fixed-duration
integration intervals of $\gls{ng:Tint;flat} = 50$ ($6250$ time steps).
As mentioned in \S~\ref{sec:meth/els/int}, the time step normalized with the
velocity of the moving frame of the $s$-th iteration, \gls{no:dt_s;flat}, is 
updated according to equation \eqref{eq:meth/dt}.
%

We initialize the problem with both fluid and particle at rest
\begin{alignat}{2}
\gls{fng:u} &=\left(0,0,0\right) \Leftrightarrow
   \gls{fno:u} && = \left(0,0,1\right)\,, \\
\gls{png:u} &=\left(0,0,0\right) \Leftrightarrow
   \gls{pno:u} && = \left(0,0,1\right)\,.
\end{alignat}
The iteration procedure on \gls{Reo} is initialized using a point-particle model
in which the drag coefficient is given by
%
\begin{equation}\label{eq:drag_model}
C_D=f(\gls{Rep})\,\frac{24}{\gls{Rep}},
\end{equation}
where $f(\gls{Rep})=1+0.15\gls{Rep}^{0.672}$ is the  factor proposed by Schiller and
Naumann \citep[][their equation 4.15]{crowe:2011}, which is reasonably good for
Reynolds number up to $800$ ($f=1$ recovers Stokes drag).
After some algebra and setting the drag force equal to the submerged weight of a sphere,
one obtains
\begin{equation}\label{eq:drag_model_update}
\gls{Rep}\,f\left(\gls{Rep}\right) = \frac{\gls{Ga;sq}}{18}\,.
\end{equation}
The initial Reynolds number is set to the
numerically determined
solution of this equation, 
resulting in $\gls{Reo_0}=145.5$ for the present case.

Table \ref{tab:example_steady_vertical__params} shows the 
 Reynolds number and the time step used to integrate equations
\eqsMov{} in every iteration.
It can be seen that the Reynolds number converges steadily to the
value $\gls{Reo}=139.97$.
This value is then the computed particle Reynolds number \gls{Rep},
which only differs from the value of the point-particle model by about 4\%.
For the sake of simplicity, iterations and integration intervals are analogous.
The value of \gls{RoG} is updated according to $\gls{RoG_sp1}=
-\left\langle\gls{png_s:u_z;flat} \right\rangle_t$ (see equation
\eqref{eq:meth/els/corr/RoG_update}), but the time window to compute the
averaged vertical velocity is now dynamically updated
\begin{equation}\label{eq:meth/example_single/avg}
\left\langle \gls{png:u_z} \right\rangle_t^{(s)} =
\cfrac{1}{\gls{tB_s}/2} %
   \int_{\gls{tB_s}/2}^{\gls{tB_s}} \gls{png:u_z}(\tau) \wrt{} \tau \,,
\end{equation}
which implies that we used the last half of the simulated time to compute
the next iterate of $\gls{Reo_sp1}=\gls{RoG_sp1}\gls{Ga}$.
During the initial transient, accelerations are large and the shorter the 
averaging window, the better the algorithm will compensate large drifts in the
computational domain.
As the case evolves, and approaches convergence, the averaging window increases,
resulting in more  stable predictions for next iterates.
This is particularly useful to handle cases in which the particle starts from
rest and shows unsteadiness in the converged state.
%

%

\begin{table} 
\caption{Integration parameters \gls{Reo} and \gls{no:dt;flat} corresponding
to each iteration of the single particle settling in the steady vertical regime
($\gls{Ga}=121, \gls{rhor}=1.5$).
\label{tab:example_steady_vertical__params}} 
\makebox[\textwidth][c]{ 
\begin {tabular}{ccc}%
\toprule Iteration ($s$)&$\gls {Reo_s}$&$\gls {no:dt_s}$\\\toprule %
\rowcolor [gray]{0.9}\ensuremath {0}&\ensuremath {145.500}&\ensuremath {9.6198\cdot 10^{-3}}\\%
\ensuremath {1}&\ensuremath {138.856}&\ensuremath {9.1805\cdot 10^{-3}}\\%
\rowcolor [gray]{0.9}\ensuremath {2}&\ensuremath {140.120}&\ensuremath {9.2642\cdot 10^{-3}}\\%
\ensuremath {3}&\ensuremath {139.940}&\ensuremath {9.2522\cdot 10^{-3}}\\%
\rowcolor [gray]{0.9}\ensuremath {4}&\ensuremath {139.973}&\ensuremath {9.2544\cdot 10^{-3}}\\%
\ensuremath {5}&\ensuremath {139.967}&\ensuremath {9.2540\cdot 10^{-3}}\\\toprule %
\end {tabular}%
 
}
\end{table}

\begin{figure} 
\makebox[\textwidth][c]{ 
\includegraphics[scale=0.8]{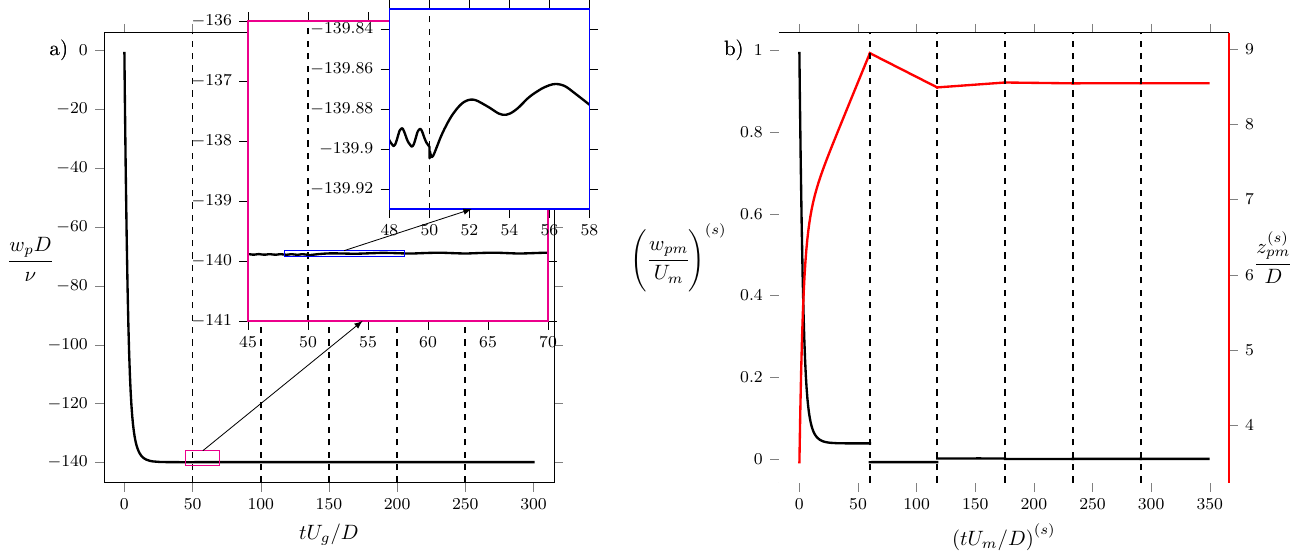}
}
\caption{Time evolution of a) vertical velocity expressed as a particle Reynolds
number (laboratory frame) and b) vertical velocity (black) and position (red) in the moving
frame corresponding to each iteration. Insets in a) are zooms of the same quantities.
\label{fig:example_steady_vertical__evol}} 
\end{figure}

Figure \ref{fig:example_steady_vertical__evol} shows the evolution of the
vertical velocity of the particle in the laboratory frame normalized with the
fluid viscosity, equivalent to an instantaneous particle Reynolds number (panel a)
and in the moving frames for every iteration (panel b).
The insets include an amplified view of the transition from one iteration
to the next one, showing a very smooth update and 
small oscillations with respect to the averaged value.
It should be noted that, although strictly speaking the converged state of this
case is stationary, the numerical approximation can show small oscillations
when using \glspl{ibm} (as we do in this example). This is a consequence of the relative
motion between the fluid and particle discretization inherent of \glspl{ibm}
\citep[see section 5.1.2 of][]{uhlmann:2005}.
In the moving frame, panel b, it can be seen how the initial rest condition of the
particle ($\gls{p:u_z} = 0$) in the laboratory frame corresponds to an upward drift in the moving frame
($\gls{pno:u_z;flat}=1$), that results in a vertical displacement of the 
particle inside the computational domain of approximately $5\gls{p:D}$.
The vertical motion in the computational domain (moving frame) is negligible
from the fourth iteration.

\section{Settling of multiple particles}
\label{sec:example_many}

\subsection{Problem description and computational setup}
\label{sec:example_many/problem}

To demonstrate the capabilities of the proposed method, we now extend the
analysis from a single particle to a suspension of many spherical particles,
\MMG{in the dilute regime. 
As in previous studies for dilute systems \citep{uhlmann:2014a,moriche:2023},
for the contact modeling we use the method proposed by \cite{glowinski1999}}.
This example highlights the main advantage of the present approach, 
its efficiency and robustness in handling multi-particle systems.

We focus on a configuration for which collective effects have been reported in
the literature. 
Specifically, we select the parameters of case M178 from \cite{uhlmann:2014a}:
$\gls{Ga}=178$, $\gls{rhor}=1.5$ and solid volume fraction
$\gls{SVF;3p}=5\cdot 10^{-3}$, corresponding to a dilute regime.
For these flow parameters, a single particle follows
a steady oblique trajectory; this is known as 
the
steady oblique regime \citep{uhlmann:2014b}.
In the triply periodic computations performed by \cite{uhlmann:2014a}, the solid
volume fraction was defined as
\begin{equation} \label{eq:SVF3p}
\gls{SVF;3p} = \frac{N_p\gls{p:vol}}{L_x\,L_y\,L_z}\,,
\end{equation}
where $N_p$ is the number of particles and $L_x$, $L_y$ and $L_z$ are the
dimensions of the computational domain.
The superscript $^*$  highlights that this definition applies to triply periodic
configurations.

\begin{figure} 
\makebox[\textwidth][c]{ 
\includegraphics[scale=0.8]{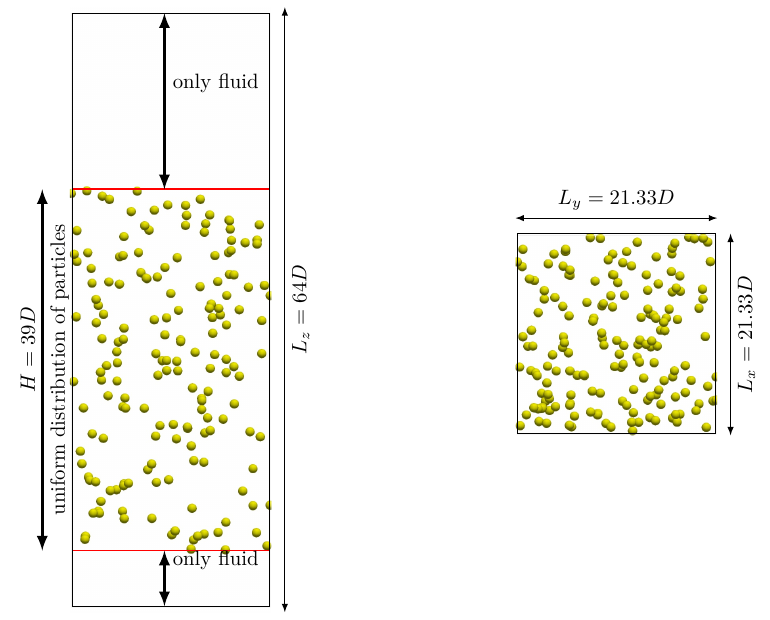}
}
\caption{a) Lateral and b) top view of the computational domain and the initial
position of the particles. Sketch-like annotations are included for clarity, but
both the representation of the computational domain and the rendering of the
particles are shown to scale.
\label{fig:example_many__init}} 
\end{figure}

In the present work, we employ a smaller computational domain than
\cite{uhlmann:2014a}, $(L_x\,L_y\,L_z)=[64/3\times64/3\times64]\gls{p:D;cb}$,
to reduce the computational cost.
Recall that the aim here is to show the feasibility of the method.
The spatial resolution is $\gls{p:D}/\gls{dx}=24$ in every direction, resulting
in a number of grid cells equal to $[512\times512\times1536]$.
The initial condition consists of a uniform distribution of particles within 
a subregion of the computational domain of height  $H=39D$; the regions above
and below are filled with fluid only (see Fig. \ref{fig:example_many__init}).
The particles are  distributed following a random Poisson process.
%

Replacing $L_z$ by $H$ in equation \eqref{eq:SVF3p}, and using the same value of
the volume fraction as \cite{uhlmann:2014a},
\begin{equation} \label{eq:SVF_io}
\gls{SVF} = \frac{N_p\gls{p:vol}}{L_x\,L_y\,H}=\,5\cdot 10^{-3},
\end{equation}
we can solve for $N_p$ and obtain $N_p=169$.
Note, that we have discarded the superscript $^*$. This is because in the
inflow–outflow configuration, the initially uniform particle concentration
becomes non-uniform as the simulation progresses. 
It develops a spatial variation that depends on the vertical coordinate, 
$\gls{svf}(z)$, as will be shown below.
%

Finally, we use a time step of $\gls{ng:dt;flat}=8\cdot10^{-3}$ and
variable-size integration intervals of minimum $200$ steps and, as mentioned in
\S~\ref{sec:meth/els/int}, the time step normalized with the velocity of the
moving frame will be updated according to equation \eqref{eq:meth/dt}.
%

\subsection{Correction algorithm}
\label{sec:example_many/corr}

In the multi-particle case, the correction algorithm requires more statistical information than for a single particle. We monitor four quantities:
the mean vertical position of the particle ensemble, its standard deviation, 
the lowest particle position and the highest particle position.
The switch from velocities to positions allows us to account 
for the spatial extent of the particle ensemble, 
which determines when corrections to the Reynolds number \gls{Reo} should be applied.
In the single-particle case, this was not needed due to the short transients involved.
We still require the mean settling velocity to correct \gls{Reo},
but this is easily obtained from the time derivative of the ensemble-averaged position
\gls{p:x_z;avg}.
The sets of reduced statistics are now 
$\gls{lab:stats} = \left\{\gls{pn:x_z;avg;flat},
                          \gls{pn:x_z;min;flat},
                          \gls{pn:x_z;max;flat},
                          \gls{pn:x_z;std;flat}
\right\}$, and
$\gls{tra:stats} = \left\{\gls{pno:x_z;avg;flat},
                          \gls{pno:x_z;min;flat},
                          \gls{pno:x_z;max;flat},
                          \gls{pno:x_z;std;flat}
\right\}$.

Each particle position in the laboratory frame is given by
\begin{equation}
\gls{p:x_z_i} = \gls{zbot} + \gls{pm:x_z_i} \,, 
\end{equation}
where \gls{zbot} is the position of the lower boundary of the computational domain
in the laboratory frame (see Appendix~\ref{app:time_zbot} for its expression).
Then, the four quantities which appear in \gls{lab:stats} are defined as
%
%
\begin{align}
\gls{p:x_z;avg} & = \frac{ \sum_{N_p} \gls{p:x_z_i}  }{\gls{Np}} \,, \\
\gls{p:x_z;min} & = \text{min}\left(\gls{p:x_z_i}\right) \,, \\
\gls{p:x_z;max} & = \text{max}\left(\gls{p:x_z_i}\right)  \,, \\
\gls{p:x_z;std} & = 
\sqrt{
\frac{ \sum_{N_p} \left(\gls{p:x_z_i}-\gls{p:x_z;avg}\right)^2 }{\gls{Np}}
}\,.
\end{align}
Thus, the mean, minimum and maximum position in both frames differ
only by the shift \gls{zbot}, whereas the standard deviation remains 
identical.
%
%
%
%
%

\begin{figure} [h!]
\makebox[\textwidth][c]{ 
\includegraphics[scale=0.8]{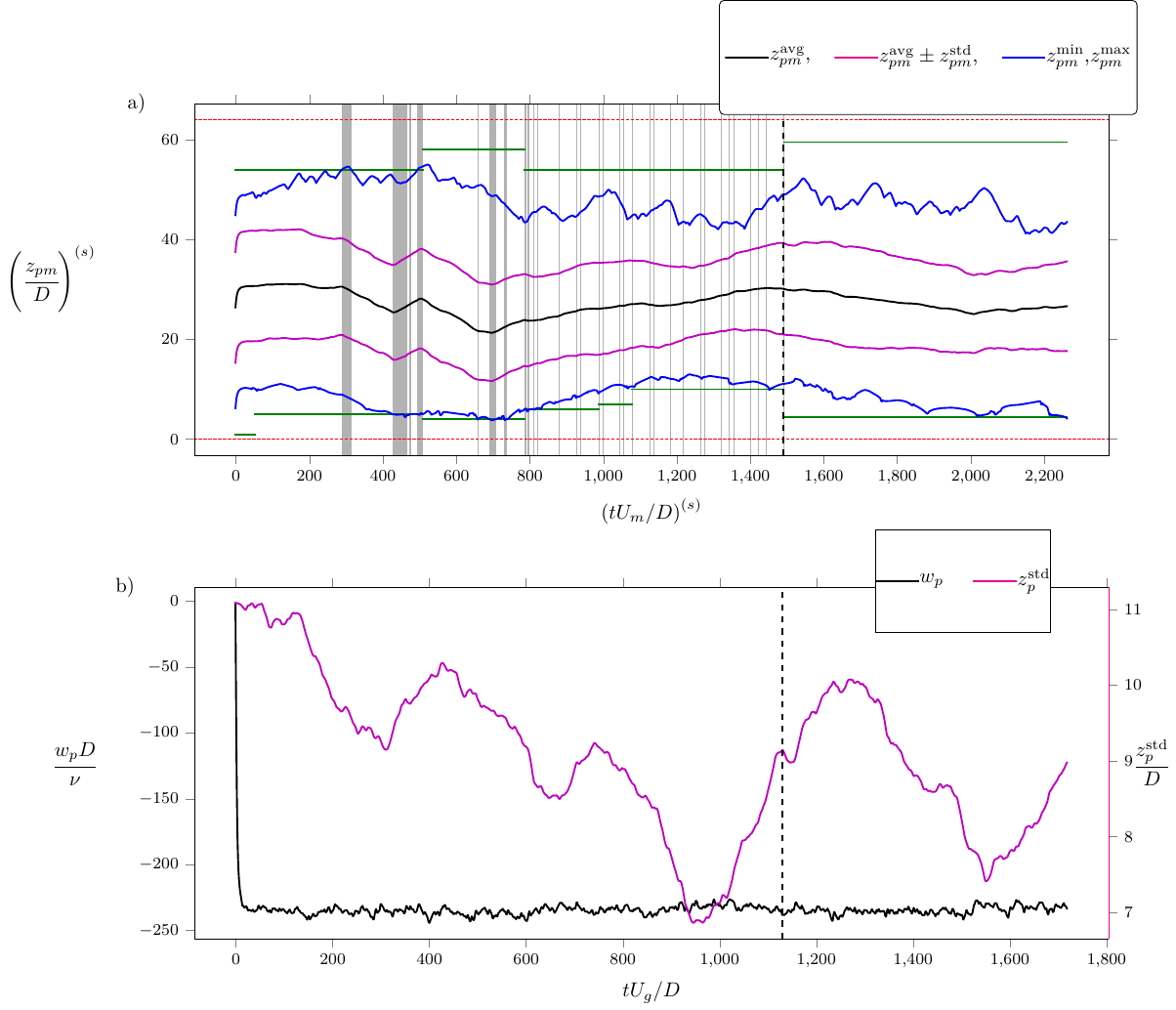}
}
\caption{Time history of a) \gls{tra:stats_s} as they are computed (moving
frame) and b) vertical velocity in the laboratory frame and width of the
distribution.
In a) the thin gray lines indicate change of moving frame, the black dashed
line the last correction, the red dashed lines lines the limits of the
computational domain and the green lines the minimum and maximum thresholds
used in the correction algorithm. 
The vertical dash line in both panels indicates the beginning of the
converged part of the simulation, in which no correction is applied, which
corresponds to $\left(t\,\gls{Uo}/D\right)^{(s)}=1489.9$ and
$t\,\gls{Ug}/D=1129.9$.
\label{fig:example_many__evol}} 
\end{figure} 

Figures \ref{fig:example_many__evol}a and b show the evolution of the metrics
during the simulation.
The correction algorithm includes the following features:
\begin{itemize}
\item If \gls{p:x_z;min} or \gls{p:x_z;max}
crosses a predefined lower or upper limit, the simulation is paused
and a new estimate of \gls{RoG} is computed.
\item  A threshold on the mean vertical 
velocity is also enforced. If exceeded, the simulation is paused
and a new estimate of \gls{RoG} is computed.
\item A vertical drift, prescribed by the user, may be
imposed to ensure the particles remain well inside the computational 
domain. For this, the equation for the next iteration is modified
with respect to \eqref{eq:meth/els/corr/RoG_update} as follows
\begin{equation}\label{eq:example_many/RoG_update}
\gls{RoG_sp1} = \cfrac{\left\langle\gls{png:u_z;flat}\right\rangle_t}{%
L_\mathrm{drift}/T_\mathrm{drift}-1} 
\end{equation}
where $L_{\mathrm{drift}}$ and $T_\mathrm{drift}$ 
denote the desired vertical drift length and drift duration, respectively, for 
the next iteration. 
These are target quantities that may not be exactly achieved during the 
subsequent time evolution. 
The closer the system is to a statistically steady state, characterized by 
reduced unsteadiness in the mean settling velocity and by being further from 
the initial transient, the more accurately these desired quantities will be 
reproduced in the next iteration.

\item The thresholds are dynamically updated based on the observed behavior
of the system.
\end{itemize}

The goal of simulating the settling of many particles over a long 
period was successfully achieved. 
Figure \ref{fig:example_many__evol} shows the entire time history 
of the simulation, including the long initial transient during 
which corrections were applied and in which the thresholds were 
gradually updated, as indicated by the various horizontal lines. 
The total interval without corrections spans approximately $600\,D/\gls{Ug}$. 
Although only this portion is used for analysis, the complete time history remains useful for feeding the correction algorithm and examining the system’s overall behavior. 

In the following section, time-averaged quantities are computed over the window $1129.9 < t\gls{Ug}/D < 1718.4$, where no further corrections are applied and the solution exhibits a reasonable degree of statistical stationarity.
During this time interval, the mean vertical position of the particle
ensemble presents a tiny drift. As a consequence,
the  Reynolds number imposed during this last interval,
$\gls{Reo}=233.52$,  is slightly different from the actual particle Reynolds number
 $\gls{Rep}=234.4$.
This difference is however smaller than $0.4\%$.
The success of the method is reflected in how small this difference is, which allows
for long time integration intervals.
This is the case even when the
 particle distribution size
shows 
oscillatory behavior 
of moderate amplitude ($3D$) and large time-scale ($300D/\gls{Ug}$),
as seen in the evolution of \gls{p:x_z;std} in figure
\ref{fig:example_many__evol}b.

\subsection{Results}
\label{sec:example_many/results}

\begin{figure} 
\makebox[\textwidth][c]{ 
\includegraphics[scale=2.0]{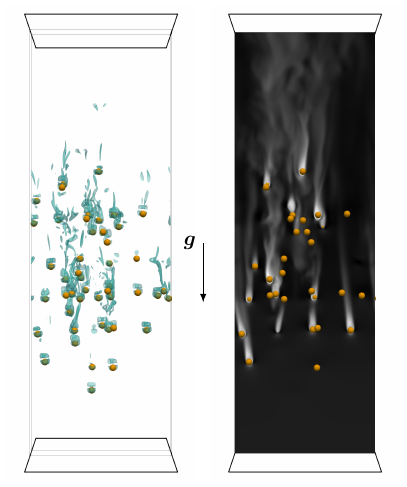}
}
\caption{Visualization of a slab of the problem of a) the Q-criterion
and b) the vertical velocity.
\label{fig:example_many__visu}} 
\end{figure}

Now let us briefly discuss some of the results that are obtained in this
new configuration. 
%
First, for illustration, 
we show a 3D visualization of a slab of the domain, representing the
particles and the vortex structure identified using the second-invariant
of the velocity gradient tensor, $Q$, (figure
\ref{fig:example_many__visu}a) and the  vertical
velocity of the fluid phase (figure \ref{fig:example_many__visu}b).
It can be seen how the wake around the particles located at the bottom part of
the distribution show a toroidal vortex (see panel a), and a clearly
tilted, straight and elongated wake (see panel b), resembling the solution of an
isolated particle in the steady oblique regime.
This is the regime that corresponds to the selected  Galileo number and 
density ratio for an isolated particle \citep{jenny:2004,uhlmann:2014b}, as mentioned above. 
%
%
%
%
Differently, particles located downstream of the first front of particles
show unsteady wakes with the characteristic  hairpin-like vortex structures,
whose formation is triggered by the hydrodynamic interactions
with the surrounding particles.

Another  phenomenon that can be examined in  this  
configuration is the perturbed flow field that develops within and 
downstream of the particle swarm.
For example, in the region of fluid surrounding 
the zone of higher particle concentration located near the center
of the domain, the downward velocity perturbations reduce
the drag experienced by the particles located further downstream.
In contrast, regions with few or no particles are also slightly perturbed (although not visible in the 
figure): because of conservation of mass, a weak upward flow develops that forms  a vertically-aligned region of increased drag.

\begin{figure} 
\makebox[\textwidth][c]{ 
\includegraphics[scale=0.8]{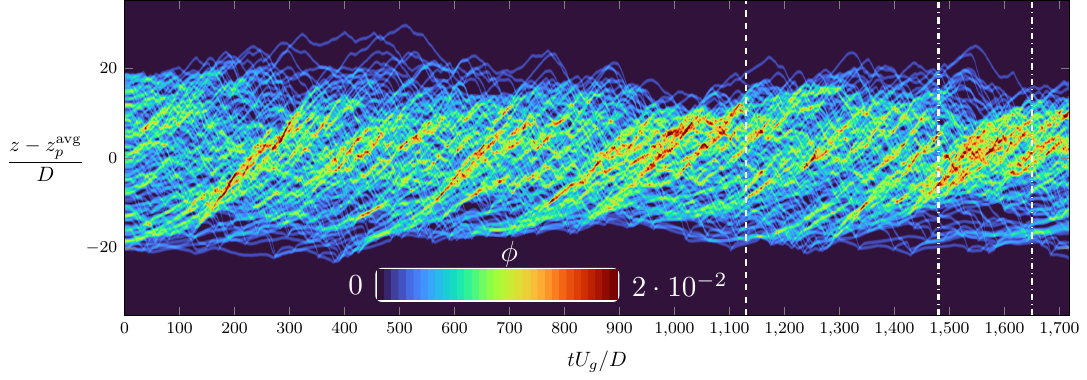}
}
\caption{Contour plot of particle concentration 
as a function of time and the vertical coordinate. The vertical dash line indicates the beginning of the
converged part of the simulation, and the dash-dotted lines are auxiliary lines
($t\gls{Ug}/D=1480,1650$) also represented in figure
\ref{fig:example_many__converged_evol_phi}a to support the discussion.
\label{fig:example_many__phi}} 
\end{figure}

\begin{figure} 
\makebox[\textwidth][c]{ 
\includegraphics[scale=0.8]{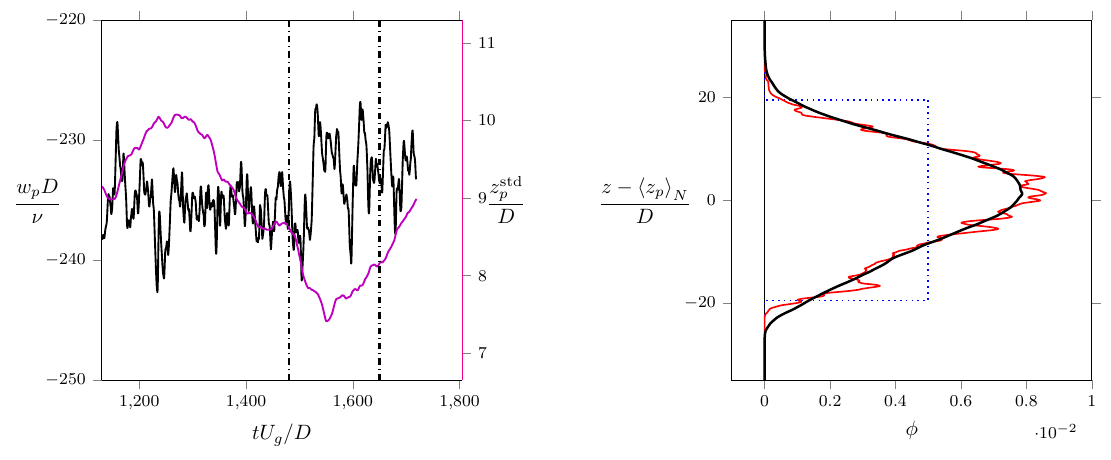}
}
\caption{a) Vertical velocity of the particle ensemble (black line)  and standard deviation of particles'
vertical position (magenta line) as a function of time.
b) Solid volume fraction averaged in the lateral directions and time
($\left\langle\gls{svf}\right\rangle_t$, red curve) and box-averaged with a
filter width of $10D$ ($\left\langle\gls{svf;z10}\right\rangle_t$, black curve).
In a) the vertical dash-dotted lines are auxiliary lines
($t\gls{Ug}/D=1480,1650$) also represented in figure
\ref{fig:example_many__phi}) to support the discussion.
In b) the initial particle concentration distribution is represented with a blue dotted line.
\label{fig:example_many__converged_evol_phi}} 
\end{figure}

Figure \ref{fig:example_many__phi} shows a contour plot of 
the particle concentration as a function of time and 
of the vertical coordinate, $\left( z-\gls{p:x_z;avg}\right)/D$.
The plot includes the whole time evolution for illustration,
but we analyse only the time interval 
where no corrections are applied, beyond the vertical 
dashed line in the figure. 
The figure shows that there are time intervals 
during which the particle ensemble is more compressed, indicated by the red patches in the figure. 
One such area is located between the two auxiliary 
lines towards the end of the simulation, $t\gls{Ug}/D=[1480,1650]$. 
At earlier times, $t\gls{Ug}/D<1480$, the particle ensemble
occupies a broader area. 
This is clearly visible in Figure \ref{fig:example_many__converged_evol_phi}a that
shows the time evolution of a measure of the
width of the particle distribution (\gls{p:x_z;std}).
This quantity presents an oscillatory behaviour with maximum and minimum
at times which are consistent
with Figure \ref{fig:example_many__phi}.

Figure \ref{fig:example_many__converged_evol_phi}a shows also the evolution of
the vertical velocity of the particle ensemble.
With the given non-dimensionalization, the quantity can be understood as
an instantaneous particle Reynolds number. 
The time evolution of this quantity presents high frequency oscillations,
probably due to the small amount of particles considered here.
Interestingly, in the interval where the particle ensemble is more compressed, the
vertical velocity of the particle ensemble is somewhat lower.
This reduction in velocity may be attributed to a hindrance effect: when particles are more 
closely packed, the fluid displaced by their motion generates upward disturbances that interfere 
with neighboring particles, thereby reducing the overall settling speed of the ensemble.
However, the number of particles considered in this study is too small for statistical
convergence and these observations need to be confirmed in simulations with 
a larger number of particles.

Finally, Figure \ref{fig:example_many__converged_evol_phi}b shows the vertical profile of the mean 
particle concentration, averaged over the time interval without corrections, together with the initial 
distribution for comparison.
Because of the small number of particles in this dilute regime, and since statistics are computed 
separately at each height, the mean concentration profile exhibits poor convergence.
To obtain a smoother representation, the figure also includes
a smoothed version, \gls{svf;z10},
obtained by applying a box-filter with a filter width of $10D$.
The results indicate that the particle distribution tends to adopt a Gaussian-like shape, with a 
maximum concentration near the center and two nearly symmetric tails extending toward the top and 
bottom.
In future simulations with a larger number of particles, it will be interesting to examine whether 
this Gaussian shape persists, how the peak concentration compares with the initial one, and how far 
the distribution tails extend beyond the initial profile, among other questions.

%
%
%
%
%
%
%

\section{Conclusions}
\label{sec:conclusions}

In this work, we advance the simulation of particle suspensions settling under gravity.
The most common setup for this type of problem is a triply periodic 
computational domain. 
However, such configurations suffer from strong vertical correlations, which 
hinder the study of phenomena such as cluster dynamics.
Here, we show that it is possible to remove the vertical periodicity and 
propose a methodology to overcome the associated numerical challenges.
%
%

With the aid of two examples of increasing complexity--a single particle
in the steady vertical regime and a many-particle case exhibiting collective
effects--we have demonstrated the robustness and efficiency of the method.
We measure the success and feasibility of the method through two aspects:
First, the approach is largely systematic; 
apart from the correcting algorithm, which is 
problem dependent, all components of the 
method are simple and well structured.
%
Furthermore, some indications have been given to deal with the openness of the
correcting algorithm.
The second aspect that demonstrates the feasibility of the method is that a
converged, free-of-corrections interval of approximately $600 \gls{p:D}/\gls{Ug}$
has been obtained in a many-particle case.
This constitutes the first simulation of this kind presented to date (to the best
 of our knowledge).

The effectiveness of the method follows from two aspects.
Even simulations initiated with coarse estimates of the moving-frame 
Reynolds number provide valuable information on the system’s settling 
dynamics. 
Each iteration refines these estimates, progressively approaching the 
converged state.
In addition, the consistent update of flow and particle variables during 
reference-frame adjustments minimizes discontinuities in the solution when 
simulations are restarted.

%
%
%
%

The proven feasibility of the method opens new possibilities in the
analysis of particle-laden flows.
By eliminating vertical periodicity, the long-term dynamics of clusters can 
now be examined under physically uncorrelated conditions.
A further observation is that the final particle concentration profile 
emerges as a result of the simulation rather than as an imposed input.
The relevant control parameter becomes the total particle weight per unit 
area, yielding a more realistic configuration in which the suspension 
naturally compresses or expands according to the underlying hydrodynamic 
interactions.

This framework also allows investigation of the influence of still fluid below the settling region on the lower particle layers.
For example, in this work we find that the enhancement of the settling velocity
is significantly lower compared to the triply periodic case.
This observation raises important questions about whether an unperturbed 
fluid layer acts effectively as a wall that suppresses collective effects, 
or whether the result reflects finite-domain limitations, among other 
possibilities.
The approach also permits detailed examination of the fluctuating flow 
induced by particle interactions: whether clusters can grow to scales large 
enough to excite broad flow structures and recover canonical turbulence, or 
whether the energy injected by particle wakes remains confined to pseudo-
turbulent regimes.%

%
%
Finally, we emphasize that the proposed methodology can be implemented in 
any numerical solver that integrates the Navier–Stokes equations in an 
inertial reference frame.
Since this includes the majority of current  codes, the present method  
broadens the accessibility of this class of problems to the wider research 
community.

%
%
%

\section{Funding}

This work was supported by the German Research Foundation (DFG)
under Project  UH 242/11$-$1 and by the Austrian Science Fund (FWF)
10.5576/PAT1793625.
For open access purposes, the author has applied a CC BY public copyright license to
any author-accepted manuscript version arising from this submission.

\bibliographystyle{jfm}
\bibliography{References}

@article{kajishima:2002,
title={Interaction between particle clusters and particle--induced turbulence},
author={Kajishima, T. and Takiguchi, S.},
journal={Int. J. Heat Fluid Flow},
volume = "23",
number = "5",
pages = "639 - 646",
year = "2002",
doi = "10.1016/S0142-727X(02)00159-5",
}

@article{jenny:2004,
  title={Instabilities and transition of a sphere falling or ascending freely in a {N}ewtonian fluid},
  author={Jenny, M. and Du{\v{s}}ek, J. and Bouchet, G.},
  journal={J. Fluid Mech.},
  volume={508},
  pages={201--239},
  year={2004}
}

@article{uhlmann:2005,
  title = "An immersed boundary method with direct forcing for the simulation of particulate flows",
  author={Uhlmann, M.},
  journal={J. Comput. Phys.},
  volume = "209",
  number = "2",
  pages = "448 - 476",
  year = "2005",
  issn = "0021-9991",
  doi = "10.1016/j.jcp.2005.03.017",
  url = "http://www.sciencedirect.com/science/article/pii/S0021999105001385"
}

@book{crowe:2011,
  title={Multiphase flows with droplets and particles},
  author={Crowe, C. T. and Schwarzkopf, J. D. and Sommerfeld, M. and Tsuji, Y.},
  year={2011},
  publisher={CRC press}
}

@article{uhlmann:2014a,
 author={Uhlmann, M. and Doychev, T.},
 journal={J. Fluid Mech.},
 title={Sedimentation of a dilute suspension of rigid spheres at intermediate {G}alileo numbers: the effect of clustering upon the particle motion},
 volume={752},
 DOI={10.1017/jfm.2014.330},
 year={2014},
 pages={310-348}
}

@article{uhlmann:2014b,
  title={The motion of a single heavy sphere in ambient fluid: a benchmark for interface--resolved particulate flow simulations with significant relative velocities},
  author={Uhlmann, M. and Du{\v{s}}ek, J.},
  journal={Int. J. Multiph. Flow},
volume = "59",
pages = "221 - 243",
year = "2014",
issn = "0301-9322",
doi = "https://doi.org/10.1016/j.ijmultiphaseflow.2013.10.010",
url = "http://www.sciencedirect.com/science/article/pii/S0301932213001559"
}

@phdthesis{doychev:2015,
    author       = {Doychev, T.},
    year         = {2015},
    title        = {The dynamics of finite--size settling particles},
    doi          = {10.5445/KSP/1000044723},
    publisher    = {{KIT Sci. Publ.}},
    isbn         = {978-3-7315-0307-1},
    issn         = {1439-4111},
    series       = {Dissertationsreihe am Institut f{\"{u}}r Hydromechanik, Karlsruher Institut f{\"{u}}r Technologie / Karlsruher Institut f{\"{u}}r Technologie (KIT), Institut f{\"{u}}r Hydromechanik},
    pagetotal    = {238},
    language     = {english},
    volume       = {2015},
    number       = {1}
}

@article{ardekani:2016,
  title={Numerical study of the sedimentation of spheroidal particles},
  author={Ardekani, M. N. and Costa, P. and Breugem, W. P. and Brandt, L.},
  journal={Int. J. Multiph. Flow},
  volume={87},
  pages={16--34},
  year={2016}
}

@article{detullio:2016,
title = {A moving--least--squares immersed boundary method for simulating the fluid–-structure interaction of elastic bodies with arbitrary thickness},
journal = {J. Comput. Phys.},
volume = {325},
pages = {201-225},
year = {2016},
issn = {0021-9991},
doi = {https://doi.org/10.1016/j.jcp.2016.08.020},
url = {https://www.sciencedirect.com/science/article/pii/S0021999116303692},
author = {M.D. {de Tullio} and G. Pascazio}
}

@article{fornari:2016,
 title={Sedimentation of finite-size spheres in quiescent and turbulent environments},
 volume={788},
 DOI={10.1017/jfm.2015.698},
 journal={J. Fluid Mech.}, 
 author={Fornari, W. and Picano, F. and Brandt, L.}, 
 year={2016},
 pages={640–669}
}

@article{huisman:2016,
  title = {Columnar structure formation of a dilute suspension of settling spherical particles in a quiescent fluid},
  author = {Huisman, S. G.. and Barois, T. and Bourgoin, M. and Chouippe, A. and Doychev, T. and Huck, P. and Morales, C. E. B. and Uhlmann, M. and Volk, R.},
  journal = {Phys. Rev. Fluids},
  volume = {1},
  issue = {7},
  pages = {074204},
  year = {2016},
  doi = {10.1103/PhysRevFluids.1.074204},
}

@article{fornari:2018,
  title={Clustering and increased settling speed of oblate particles at finite {R}eynolds number},
  author={Fornari, W. and Ardekani, M. N. and Brandt, L.},
  journal={J. Fluid Mech.},
  volume={848},
  pages={696--721},
  year={2018}
}

@InCollection{uhlmann:2023,
  author = 	 {M. Uhlmann and J. Derksen and A. Wachs and L.-P. Wang and M. Moriche},
  title = 	 {Efficient methods for particle-resolved direct numerical simulation},
  booktitle = 	 {Modelling approaches and computational methods for particle-laden turbulent flows},
  publisher = {Academic press},
  editor = 	 {S. Subramaniam and S. Balachandar},
  year = 	 2023,
  pages = 	 {147-184},
}

@book{subramaniam:2023,
  title={Modeling approaches and computational methods for particle-laden turbulent flows},
  editor={Subramaniam, S. and Balachandar, S.},
  year={2023},
  publisher={Academic Press}
}

@article{moriche:2021,
  title={A single oblate spheroid settling in unbounded ambient fluid: a benchmark for simulations in steady and unsteady wake regimes},
  author={Moriche, M. and Uhlmann, M. and Du{\v{s}}ek, J.},
  journal={Int. J. Multiph. Flow},
  volume={136},
  pages={103519},
  year={2021},
  doi={10.1016/j.ijmultiphaseflow.2020.103519}
}

@article{seyed-ahmadi:2021,
  title = {Sedimentation of inertial monodisperse suspensions of cubes and spheres},
  author = {Seyed-Ahmadi, A. and Wachs, A.},
  journal = {Phys. Rev. Fluids},
  volume = {6},
  issue = {4},
  pages = {044306},
  year = {2021},
  doi = {10.1103/PhysRevFluids.6.044306},
}

@article{lu:2023,
  title={The dynamics of suspensions of prolate spheroidal particles -- {E}ffects of volume fraction},
  author={Lu, J. and Xu, X. and Zhong, S. and Ni, R. and Tryggvason, G.},
  journal={Int. J. Multiph. Flow},
  volume={165},
  pages={104469},
  year={2023}
}

@article{moriche:2023,
  title={On the clustering of low-aspect-ratio oblate spheroids settling in ambient fluid},
  author={Moriche, M. and Hettmann, D. and Garc{\'\i}a-Villalba, M. and Uhlmann, M.},
  journal={J. Fluid Mech.},
  volume={963},
  pages={A1},
  year={2023}
}

@article{jiang:2024,
  title={Prolate spheroids settling in a quiescent fluid: clustering, microstructures and collisions},
  author={Jiang, X. and Xu, C. and Zhao, L.},
  journal={J. Fluid Mech.},
  volume={1000},
  pages={A49},
  year={2024}
}

@article{garciavillalba:2025,
  title={Numerical methods for multiphase flows},
  author={Garcia-Villalba, M. and Colonius, T. and Desjardins, O. and Lucas, D. and Mani, A. and Marchisio, D. and Matar, O. K. and Picano, F. and Zaleski, S.},
  journal={Int. J. Multiph. Flow},
  volume={191},
  pages={105285},
  year={2025}
}

@article{villafane2025,
  title={50 years of {International Journal of Multiphase Flow: E}xperimental methods for dispersed multiphase flows},
  author={Villafa{\~n}e, L. and Aliseda, A. and Ceccio, S. and DiMarco, P. and Machicoane, N. and Heindel, T. J.},
  journal={Int. J. Multiph. Flow},
  volume={189},
  pages={105239},
  year={2025}
}

@article{marchioli2025,
  title={Particle-laden flows},
  author={Marchioli, C. and Bourgoin, M. and Coletti, F. and Fox, R. and Magnaudet, J. and Reeks, M. and Simonin, O. and Sommerfeld, M. and Toschi, F. and Wang, L.-P. and Balachandar, S.},
  journal={Int. J. Multiph. Flow},
  volume={191},
  pages={105291},
  year={2025}
}

@article{catalan2024,
  title={On the settling of a spherical particle in slightly perturbed ambient fluid},
  author={Catal{\'a}n, J. M. and Moriche, M. and Flores, O. and Garc{\'\i}a-Villalba, M.},
  journal={Acta Mech.},
  volume={235},
  number={4},
  pages={2479--2493},
  year={2024}
}

@article{kempe2012,
  title={Collision modelling for the interface-resolved simulation of spherical particles in viscous fluids},
  author={Kempe, T. and Fr{\"o}hlich, J.},
  journal={J. Fluid Mech.},
  volume={709},
  pages={445--489},
  year={2012}
}

@article{costa2015,
  title={Collision model for fully resolved simulations of flows laden with finite-size particles},
  author={Costa, P. and Boersma, B. J. and Westerweel, J. and Breugem, W.-P.},
  journal={Phys. Rev. E},
  volume={92},
  number={5},
  pages={053012},
  year={2015}}

@article{kidanemariam2014,
  title={Interface-resolved direct numerical simulation of the erosion of a sediment bed sheared by laminar channel flow},
  author={Kidanemariam, A. G. and Uhlmann, M.},
  journal={Int. J. Multiph. Flow},
  volume={67},
  pages={174--188},
  year={2014}
}

@article{glowinski1999,
  title={A distributed {Lagrange} multiplier/fictitious domain method for particulate flows},
  author={Glowinski, R. and Pan, T.-W. and Hesla, T. I. and Joseph, D. D.},
  journal={Int. J. Multiph. Flow},
  volume={25},
  number={5},
  pages={755--794},
  year={1999}
}

\appendix

\section{Technical details on the treatment of time}
\label{app:time_zbot}

%
Although time is treated in a simple way, the relationship between
the time normalized by gravitationally scaling,
\gls{ng:t;flat}, and the time normalized by the moving-frame velocity,
\gls{no:t;flat}, requires some discussion.
In some cases, this distinction can be ignored, 
for example, when analyzing a single particle in 
\S~\ref{sec:example_single_particle}.
However, when determining the position of the bottom boundary
of the moving domain in the laboratory frame,  
\gls{zbot}, as in \S~\ref{sec:example_many/corr} or in more
complex situations such as imposing continuous turbulence at the inlet,
 this issue must be addressed.
%
%
%
%
%
Therefore, in this appendix, we establish the relation between the time
in the $s$-th moving frame and the time in the laboratory frame.

Assuming that the origin of time ($t=0$) is the beginning of the simulation,
the times normalized by \gls{Ug} and \gls{Uo} in the first iteration ($s=0$)
are simply related by
\begin{equation}
\gls{no_0:t} = \gls{RoG_0}\gls{ng:t} \,;
\end{equation}
recall that $\gls{RoG_s}=\gls{Reo_s}/\gls{Ga}$.
Since the moving frame velocity changes from iteration to iteration, this needs to be taken into account leading to, for the $s$-th iteration, 
%
\begin{equation}\label{eq:t}
\gls{no_s:t} = \sum_{i=0}^{s-1}\left(\gls{RoG_i} \frac{\gls{T_i}\,\gls{Ug}}{\gls{p:D}}\right)
             + \gls{RoG_s}\frac{\left(t-\gls{tA_s}\right)\gls{Ug}}{D}
\,,
\end{equation}
where $\gls{T_i}=\gls{tB_i}-\gls{tA_i}$ is the interval duration of the $i$-th iteration.

The position of the lower boundary in the laboratory frame,
\gls{zbot}, 
is easily obtained since during each iteration the reference frame
travels at constant speed. 
Therefore, 
for the $s$-th iteration it is given by 
\begin{equation}
\gls{n:zbot_s}(t) = -\sum_{i=0}^{s-1} \left( \frac{\gls{T_i}\gls{Uo_i}}{D}\right) 
                  -\frac{\left(t-\gls{tA_s}\right)\gls{Uo_s} }{D}
\,, 
\end{equation}
where we have assumed that $\gls{zbot}(t=0) = 0$.
Introducing in this expression, the factor 
$\gls{RoG_s}=\gls{Reo_s}/\gls{Ga}=\gls{Uo_s}/\gls{Ug}$, leads to
%
\begin{equation}
\gls{n:zbot_s}(t)
                = -\sum_{i=0}^{s-1} \left( \gls{RoG_i}\frac{\gls{T_i}\gls{Ug}}{D}\right) 
                -\gls{RoG_s}\frac{\left(t-\gls{tA_s}\gls{Ug}\right)}{D} = 
 -\gls{no_s:t} \,.
\end{equation}
In the last equality, equation \eqref{eq:t} has been used.

\end{document}